\documentclass[iop]{emulateapj}
\usepackage{amsmath}
\usepackage{natbib}

\begin{document}
\title{Turbulence-driven coronal heating and improvements to empirical forecasting of the solar wind}
\author{Lauren N. Woolsey\altaffilmark{1},
and
Steven R. Cranmer\altaffilmark{1},}
\altaffiltext{1}{Harvard-Smithsonian Center for Astrophysics,
60 Garden Street, Cambridge, MA 02138,USA}
\begin{abstract} Forecasting models of the solar wind often rely on simple parameterizations of the magnetic field that ignore the effects of the full magnetic field geometry. In this paper, we present the results of two solar wind prediction models that consider the full magnetic field profile and include the effects of Alfv\'en waves on coronal heating and wind acceleration. The one-dimensional MHD code ZEPHYR self-consistently finds solar wind solutions without the need for empirical heating functions. Another 1D code, introduced in this paper (The Efficient Modified-Parker-Equation-Solving Tool, TEMPEST), can act as a smaller, stand-alone code for use in forecasting pipelines. TEMPEST is written in Python and will become a publicly available library of functions that is easy to adapt and expand. We discuss important relations between the magnetic field profile and properties of the solar wind that can be used to independently validate prediction models. ZEPHYR provides the foundation and calibration for TEMPEST, and ultimately we will use these models to predict observations and explain space weather created by the bulk solar wind. We are able to reproduce with both models the general anticorrelation seen in comparisons of observed wind speed at 1 AU and the flux tube expansion factor. There is significantly less spread than comparing the results of the two models than between ZEPHYR and a traditional flux tube expansion relation. We suggest that the new code, TEMPEST, will become a valuable tool in the forecasting of space weather.\end{abstract}
\section{Introduction}
The solar wind is a constant presence throughout the heliosphere, affecting cometary tails, planetary atmospheres, and the interface with the interstellar medium. Identifying the acceleration mechanism(s) that power the wind remains one of the key unsolved mysteries in the field. Theorists have proposed a variety of physical processes that may be responsible, and these processes are invoked in models that seek to explain both the heating of the solar corona and the acceleration of the solar wind. Such models are often categorized by their primary use of either magnetic reconnection and the opening of closed magnetic loops (Reconnection/Loop-Opening models, RLO) or the generation of magnetoacoustic and Alfv\'en waves and the turbulence created by them (Wave/Turbulence-Driven models, WTD). Several reviews have discussed the many suggested models and the associated controversies \citep{1993SoPh..148...43Z, 1996SSRv...75..453N, 2006SoPh..234...41K, 2009LRSP....6....3C}. \\
\indent RLO models require closed field lines, where both footpoints of magnetic flux tubes are anchored to the photosphere. Interactions between neighboring closed loops or between closed and open field lines lead to magnetic reconnection, which releases stored magnetic energy when the magnetic topology is reconfigured. Reconnection in closed field regions has been suggested to play a role in streamers \citep{1999JGR...104..521E, 2011ApJ...731..112A} and in the quiet Sun on supergranular scales \citep{1992sws..coll....1A, 1999JGR...10419765F, 2003JGRA..108.1157F, 2006ApJ...642.1173S, 2011ApJ...731L..18M, 2013ApJ...770....6Y}. However, \citet{2010ApJ...720..824C} provided evidence that the complex and continuous evolution of this so-called ``magnetic carpet'' \citep{1998ASPC..154..345T} of open and closed field lines may not provide enough energy to accelerate the outflow to match in situ measurements of wind speed. \\
\indent Alternatively, WTD models are useful for explaining heating and wind acceleration in regions of the Sun where the flux tubes are primarily open, that is, they are rooted to the photosphere by only one footpoint and reconnection is less likely to release significant amounts of energy. In this case, Alfv\'en waves and magnetoacoustic oscillations can be launched at the footpoints when the flux tube is jostled by convection at the photosphere. As the density of the solar atmosphere drops with height, the waves are partially reflected; counter-propagating waves interact and generate magnetohydrodynamic (MHD) turbulence. This turbulence generates energy at large scales, and the break-up of eddies causes an energy cascade down to smaller scales where the energy can be dissipated as heat at a range of heights. WTD models have naturally produced solar winds with properties that match observed outflows in the corona and further out in the heliosphere \citep{1986JGR....91.4111H, 1991ApJ...372L..45W, 1999ApJ...523L..93M, 2006JGRA..111.6101S, 2007ApJS..171..520C,2010ApJ...708L.116V}. This paradigm for solar wind acceleration has, however, also been challenged \citep{2010ApJ...711.1044R}, so perhaps the answer to the entire question of coronal heating is more complex than previously thought.\\
\indent One of the most striking aspects of the observations of the solar wind is the appearance of a bimodal distribution of speeds at 1 AU. The existence of separate components of the outflow has been observed since {\it Mariner 2} began collecting data in interplanetary space \citep{1962Sci...138.1095N, 1966JGR....71.4469N}. The fast wind has asymptotic wind speeds above roughly 600 km s$^{-1}$ and is characterized by low densities, low variability, and photospheric abundances. The slow wind, however, has speeds at 1 AU at or below 450 km s$^{-1}$ and is chaotic, with high densities and enhanced abundances of low-FIP elements \citep{1995SSRv...72...49G}. It has been widely accepted that fast wind streams originate from coronal holes, which are characterized by unipolarity, open magnetic field, and lower densities \citep[and references therein]{1977chhs.conf.....Z, 2009LRSP....6....3C}. The location of slow wind is more of a mystery. Slow solar wind has often been attributed to sources in the streamer belt \citep{2012JGRA..117.4104C}, but recent progress suggests that perhaps pseudostreamers or the edges of coronal holes may significantly contribute to this slower population \citep{2012ApJ...749..182W, 2011ApJ...731..112A, 2004JASTP..66.1295A}. However, a variety of acceleration mechanisms have been proposed along with these suggested slow wind sources. In this paper, we investigate many of these sources and coronal structures using a single theoretical framework, allowing us to determine if the different populations of solar wind can be explained by simply a difference in their region of origin.\\
\indent Current observations have not been able to distinguish between the many competing theoretical models, as many models have a variety of free parameters that can be adjusted to fit observations without specifying all of the physics. To compare the validity of these models at different points in the solar cycle and for different magnetic field structures on the Sun, the community needs flexible tools that predict wind properties using a limited number of input parameters that are all based on observations and fundamental physics. In this project, we study the extent of magnetic field structures that can produce solar wind that matches observations using two WTD models. In Section 2, we set up a grid of flux tube models as a parameter study of a broad range of open magnetic structures throughout the solar cycle. We present in Section 3 the analysis of this grid of models using ZEPHYR \citep{2007ApJS..171..520C}. We introduce the new code TEMPEST in Section 4 and discuss its use as a forecasting tool. In Section 5 we compare the results of ZEPHYR and TEMPEST and discuss differences in the models. Finally, in Section 6 we discuss these results and their importance in solving the coronal heating problem.

\section{Variation and Dynamic Range of Magnetic Field Structures}
\begin{figure}
\includegraphics[width=0.45\textwidth]{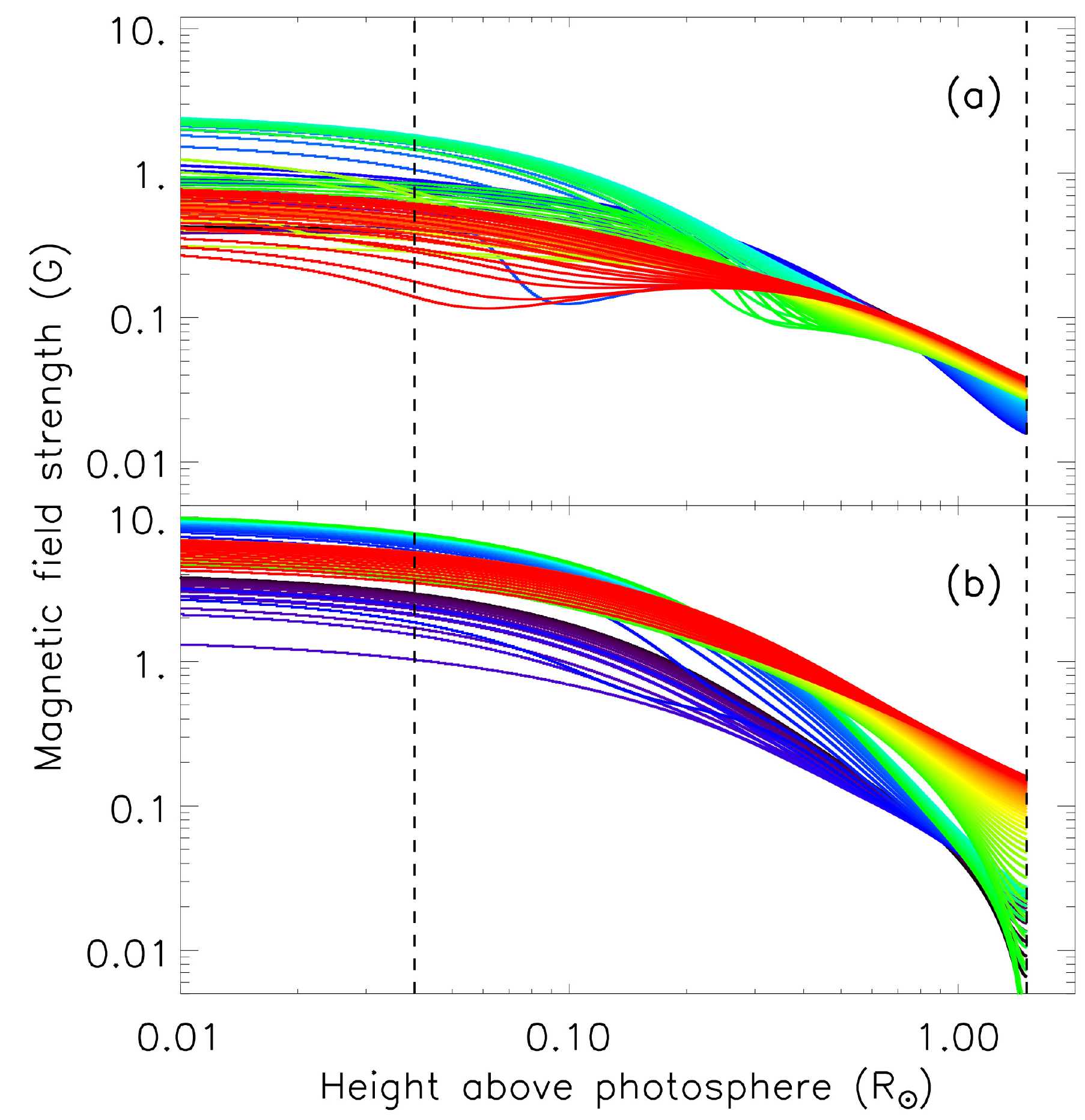}
\caption{PFSS extrapolations from WSO magnetograms for cycle 23 at (a) solar minimum in May 1996 (CR 1909) and at (b) solar maximum, in March 2000 (CR 1960). Magnetic field lines traced at the source surface along the equator, longitude denoted by color. Dashed lines show heights used in calculating the expansion factor, $z_{base} = 0.04 R_{\odot}$ and $z_{ss} = 1.5 R_{\odot}$.}
\label{fig:wso_pfss}
\end{figure}

\begin{figure}
\includegraphics[width=0.45\textwidth]{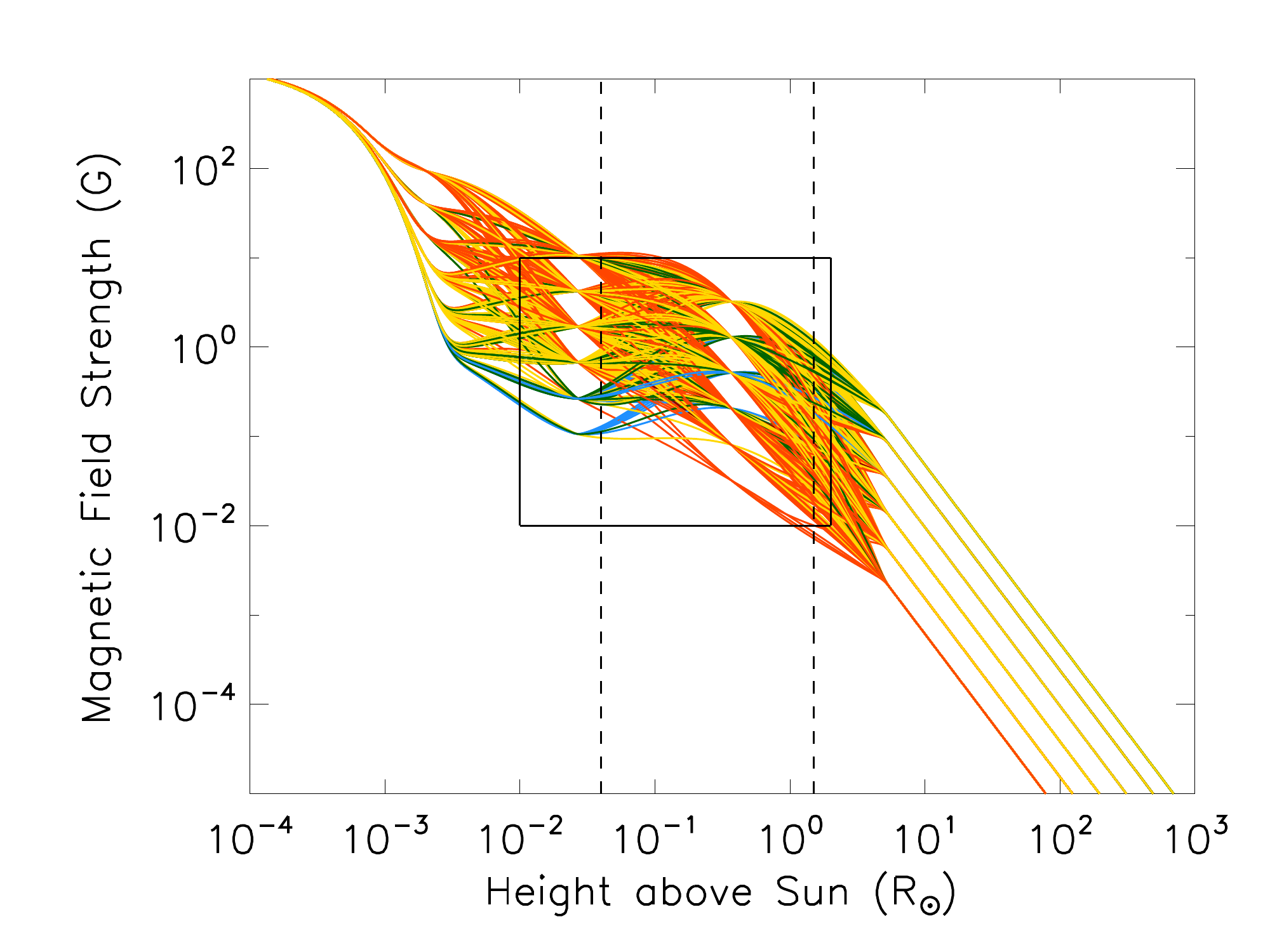}
\caption{Input flux tube models for ZEPHYR, where color indicates 200 km s$^{-1}$ wide bins of wind speed at 1 AU. Red signifies the slowest speed bin while blue shows the highest speed models (see Figure \ref{fig:tempest_temps} for key). Dashed vertical lines are plotted at the heights used for the expansion factor calculations ($z_{base} = 0.04 R_{\odot}$ and $z_{ss} = 1.5 R_{\odot}$). The black box provides the relative size of the plot ranges in Figure \ref{fig:wso_pfss} for reference.}
\label{fig:zt25}
\end{figure}

For several decades, the solar physics community has relied heavily on a single measure of the magnetic field geometry to forecast the solar wind properties at 1 AU, the so-called {\it expansion factor}. \cite{1990ApJ...355..726W} defined the expansion factor relative to the source surface radius \citep{1969SoPh....6..442S,1969SoPh....9..131A} as: \begin{equation}\label{eq:test} f_{s} = \left(\frac{R_{base}}{R_{ss}}\right)^{2} \left[\frac{B(R_{base})}{B(R_{ss})}\right] \end{equation} In Equation \eqref{eq:test}, the subscript ``base'' signifies the radius of the photospheric footpoint of a given flux tube and ``ss'' refers to the source surface, typically set to $r$ = 2.5 R$_{\odot}$. Potential Field Source Surface (PFSS) modeling assumes that ${\bf \nabla} \times {\bf B} = 0$, the source surface is a surface of zero potential, and field lines that reach this height are forced to be radial and defined as ``open'' to the heliosphere.  Using the expansion factor of Equation \eqref{eq:test}, \cite{1990ApJ...355..726W} determined an empirical relationship between $f_{s}$ and the radial outflow at 1 AU ($u_{AU}$). They binned observed expansion factors and gave typical outflow speeds at 1 AU for each bin. They found that, for $f_{s} < 3.5$, $u_{AU} \approx 700$ km s$^{-1}$ and for $f_{s} > 54$, $u_{AU} \approx 330$ km s$^{-1}$. The key point from this simple model is that the fastest wind comes from flux tubes with the lowest expansion factors and vice versa.\\
\indent The Wang-Sheeley empirical relation was used throughout the field for a decade before it was modified by \cite{2000JGR...10510465A}. They used a two-step process to make four-day advanced predictions, first defining the relation between expansion factor and wind speed {\it at the source surface} and then propagating that boundary condition of the solar wind to the radius of the Earth's orbit, including the effects of stream interactions. The initial step relies on a similar empirical fit to assign a velocity at the source surface based on the expansion factor in Equation \eqref{eq:test}, and is set by the following expression: \begin{equation}\label{eq:argepizzo} u(f_{s}) = 267.5 + \left(\frac{410}{f_{s}^{2/5}}\right) \end{equation}
\indent Because this combined Wang-Sheeley-Arge (WSA) model is often the exclusive method used for forecasting the solar wind, it is important for us to consider the efficacy of this method correctly matching observations. Early comparisons of the WSA model and observations gave correlation coefficients often at or below 0.5 for a given subset of the observations, and over the full three-year period they considered, the best method used had an overall correlation coefficient of 0.39  \citep{2000JGR...10510465A}. \cite{2005AdSpR..35.2185F} found that comparing the wind speed and expansion factor led to a correlation coefficient of 0.56. Expansion on WSA with semi-empirical modeling predicted solar maximum properties well, but produced up to 100 km s$^{-1}$ differences in comparison to observations during solar minimum \citep{2007ApJ...654L.163C}. This suggests that the community could benefit from a better prediction scheme than this simple reliance on the expansion factor. More recently, the WSA model has been used in conjunction with an ideal MHD simulation code called ENLIL (\citealp{2004JGRA..109.2116O}; see also \citealp{2011JGRA..116.3101M}). Even with the more sophisticated MHD code, it is still very difficult to make accurate predictions of the wind speed based on only a single measure of the magnetic field geometry at the Sun, $f_{s}$. Prediction errors are often attributed to the fact that these models do not account for time evolution of the synoptic magnetic field, but we also believe that the limitations of the simple WSA correlation may be to blame as well.\\
\indent There is a specific structure type observed on the Sun for which the WSA model is consistently inaccurate.
At solar minimum, the Sun's magnetic field is close to a dipole, with large polar coronal holes (PCH) where the field is open to the greater heliosphere and a belt of helmet streamers where the northern and southern hemispheres have opposite polarity radial magnetic field strength. However, if an equatorial coronal hole (ECH) is present with the same polarity of the PCH of that hemisphere, there is an additional structure that has a shape similar to a helmet streamer but has the same polarity on either side of it that fills the corona between the ECH and PCH. Early work \citep[][and references therein]{1998JGR...103.2021E, 1999SoPh..188..277E} refers to these structures as ``streamer belts without a neutral line,'' and this is the most important distinction as these structures contain no large current sheets, whereas helmet streamers nearly always are a part of the heliospheric current sheet (HCS). \cite{2007ApJ...658.1340W} coined the term ``pseudostreamers'' and discuss observations of the solar wind emanating from such structures. They found that the $v-f_{s}$ relationship vastly overestimates the wind speed from pseudostreamers because these structures are characterized by squashed expansion but produce slow wind \citep[see also][]{2012ApJ...749..182W}. Further work by \cite{2005AdSpR..35.2185F} found that comparisons using the parameter $B_{\odot}/f_{s}$, where $B_{\odot}$ is the mean photospheric magnetic field strength of the flux tube, yielded a more accurate prediction of the wind speed. Comparing this parameter to our definition of $f_{s}$ in Equation \eqref{eq:test} suggests that only the magnetic field at the source surface is needed to describe the relationship between magnetic field geometry and solar wind properties. \cite{2006ApJ...640L..75S} provides a theoretical interpretation for why this parameter ($B_{\odot}/f_{s}$) works well to describe solar wind accelerated by Alfv\'en waves.\\
\indent In order to investigate the full range of open magnetic fields that exist throughout the solar cycle, we examine PFSS extrapolations from full Carrington rotation (CR) magnetograms taken by Wilcox Solar Observatory \citep{1986SoPh..105..205H}. Figure \ref{fig:wso_pfss} shows two representative data sets from solar cycle 23. What is most important to note is that the flux tubes extrapolated from observations do not always decrease monotonically. Two flux tubes with identical values of $f_{s}$ may look significantly different at heights between the photosphere and source surface. These differences at middle heights could have a significant impact on the properties of the resulting solar wind. It is for this reason that we consider a wide array of magnetic field models in this project. The two CRs presented in Figure \ref{fig:wso_pfss} do not reflect all possible magnetic field geometries, but they provide an idea of how the magnetic field changes throughout a solar cycle. In order to investigate the entire parameter space of open magnetic geometries, we looked at the absolute maximum and minimum field strengths at several heights between the source surface ($z = 1.5 R_{\odot}$, i.e. $r = 2.5 R_{\odot}$) and a height of $z = 0.04 R_{\odot}$, the scale of supergranules which is representative of the resolution of the Wilcox magnetograms, for the previous three solar cycles. We then created a grid of models spanning strengths slightly beyond those observed from solar minimum to solar maximum.\\
\indent We also investigate specific geometries associated with the open field lines in and around structures observed in the corona such as helmet streamers and pseudostreamers. Using the standard coronal hole model of \cite{2007ApJS..171..520C} as a baseline, we specified the magnetic field strength at four set heights (z = 0.002, 0.027, 0.37, and 5.0 $R_{\odot}$) and connected these strengths using a cubic spline interpolation in the quantity $\log{B}$. Thus, we include all combinations of sets of magnetic field strengths at ``nodes'' between the chromosphere and a height beyond which flux tubes expand into the heliosphere radially such that $B \propto r^{-2}$. To account for the way in which magnetic fields are thought to trace down to the intergranular network, we add two hydrostatic terms in quadrature to the potential field at heights below $z \approx 10^{-3}$ R$_{\odot}$ \citep[see][]{2013ApJ...767..125C}. The resulting 672 models are shown in Figure \ref{fig:zt25}. They span the full range of field strengths measured at 1 AU as found in the OMNI solar wind data sets. The central 90\% of the OMNI data lie between $3 \times 10^{-6}$ and $7 \times 10^{-5}$ Gauss, and our models have magnetic field strengths at 1 AU between $10^{-6}$ and $10^{-4}$ Gauss.\\
\section{ZEPHYR Analysis}
\cite{2007ApJS..171..520C} introduced the MHD one-fluid code ZEPHYR and showed that ZEPHYR can accurately match observations of the solar wind. In that paper, the authors based their magnetic field geometry on the configuration of \cite{1998A&A...337..940B} and the modifications by \cite{2005ApJS..156..265C}. 
\begin{subequations}
The equations of mass, momentum, and internal energy conservation solved by ZEPHYR are listed below:\begin{small}
\begin{align}
	\frac{\partial \rho}{\partial t} + \frac{1}{A}\frac{\partial}{\partial r}(\rho u A) = 0\\
	\frac{\partial u}{\partial t} + u\frac{\partial u}{\partial r} + \frac{1}{\rho}\frac{\partial P}{\partial r} = -\frac{G M_{\odot}}{r^{2}} + D\\
	\frac{\partial E}{\partial t} + u\frac{\partial E}{\partial r} + \left(\frac{E + P}{A}\right)\frac{\partial}{\partial r}(uA) = Q_{rad} + Q_{cond} + Q_{A} + Q_{S}
\end{align}\end{small}
\end{subequations}
In these equations, the cross-sectional area $A$ is a stand-in for $1/B$ since magnetic flux conservation requires that, along a given flux tube, $BA$ is constant. $D$ is the bulk acceleration from wave pressure and $Q_{A}$ and $Q_{S}$ are heating rates due to Alfv\'en and sound waves. \cite{2007ApJS..171..520C} also assumed the number densities of protons and electrons are equal. This one-fluid approximation means that we are unable to include effects such as preferential ion heating, but the base properties of the solar wind produced are accurate. As ZEPHYR solves for a steady-state solution, we can neglect the time-derivatives. \cite{1977ApJ...215..942J} showed that \begin{equation}D = -\frac{1}{2\rho}\frac{\partial U_{A}}{\partial r} - \left(\frac{\gamma + 1}{2\rho}\right)\frac{\partial U_{S}}{\partial r} - \frac{U_{S}}{A\rho}\frac{\partial A}{\partial r}\end{equation}
where $U_{A}$ and $U_{S}$ are the Alfv\'enic and acoustic energy densities, respectively. With the above equations and wave action conservation, ZEPHYR uses two levels of iteration to converge on a steady-state solution for the solar wind. The only free parameters ZEPHYR requires as input are 1) the radial magnetic field profile and 2) the wave properties at the footpoint of the open flux tube. In this project, we do not change the wave properties at the photospheric base from the standard model presented by \cite{2007ApJS..171..520C}. The process of analyzing grids of models for this project led to some code fixes and we use this updated version of ZEPHYR for all work presented here \citep[see also][]{2013ApJ...767..125C}.
\subsection{Determining the Physically Signficant Critical Point}
\cite{1958ApJ...128..664P} used an isothermal, spherically symmetric corona to find a family of solutions to the hydrodynamic conservation equations, and stated that there was a single physically meaningful solution where the solar wind transitioned from subsonic at low heights to supersonic above the so-called ``critical point.'' For a considerable amount of time, there was concern in the community about the likelihood of the Sun finding this single critical solution as Parker postulated, rather than one of the many ``solar breeze'' solutions that never become supersonic. This uneasiness was put to rest when \cite{1994ApJ...432L..55V, 2001Ap&SS.277..157V} showed that this critical solution (and the opposite, Bondi accretion) is stable for steady-state solutions. Solar breeze solutions evolve to the trans-sonic critical solution after conditions that may have produced them are perturbed.\\
\indent Including a wave pressure term in the momentum equation and moving away from a spherically symmetric geometry allows the momentum conservation equation to produce multiple critical points where the wind speed reaches the local critical speed, creating an even more compicated solution topology. For this project, we have revised the method by which ZEPHYR determines the true critical point from its original version used by \cite{2007ApJS..171..520C}. The revised method is described below. We solve for heights where the right-hand side (RHS) of the equation of motion,
\begin{equation}\label{eq:parkereq}\left(u - \frac{u_{c}^{2}}{u}\right)\frac{du}{dr} = \frac{-GM_{*}}{r^{2}} - u_{c}^{2}\frac{dlnB}{dr} - a^{2}\frac{dlnT}{dr} +\frac{Q_{A}}{2\rho (u+V_{A})}\end{equation}
(rewritten from Equation (3b), neglecting sound speed terms) is equal to zero, i.e. heights where the outflow speed is equal to the critical speed, whose radial dependence is defined by \begin{equation}u_{c}^{2} = a^{2} + \frac{U_{A}}{4\rho}\left(\frac{1+3M_{A}}{1+M_{A}}\right).\label{eq:ucrit}\end{equation}
Here, $a$ is the isothermal sound speed, $a^{2} = kT/m$, where ``isothermal'' means that in the definition, $\gamma = 1$, and $M_{A}$ is the Alfv\'enic Mach number, $M_{A} = u/V_{A}$. At heights where $u(r) = u_{c}(r)$, the critical slope must have two non-imaginary values at the true critical point in order to create an X-point in the first place, and for the solar wind we take the positive slope: the range of possible topologies of critical points beyond the X-point has been presented by \cite{1977JGR....82...23H}. The correct X-point, if there is more than one, lies at the global minimum of the integrated RHS of Equation \eqref{eq:parkereq}, based on the work of \cite{1976SoPh...49...43K}. The authors showed that if there are multiple critical points but only one of which is an X-point, the outermost root of the RHS is always the location of the X-point. However, if there are multiple X-points, the global minimum of the function $F(r)$ below in Equation \eqref{eq:Frhs} is the location of the X-point that the stable wind solution must follow, i.e. \begin{equation}\label{eq:Frhs} F(r) = \int_0^r \! {\rm RHS} \, dr'.\end{equation} This was reconfirmed in the more recent paper by \cite{2003ApJ...598.1361V}.
\subsection{ZEPHYR Results for the 672-Model Grid}
We now present some of the most important relations between the solar wind properties output by ZEPHYR and the input grid of magnetic field profiles. For the number of iterations we allowed in ZEPHYR, a subset of the models converged properly to a steady-state solution (i.e. a model is considered converged if the internal energy convergence parameter $\langle \delta E \rangle$ as defined by \cite{2007ApJS..171..520C} is $\le 0.07$). We analyze only the results of these converged 428 models (out of the total 672) in the figures presented in the following subsections. Recall that the solar wind forecasting community relies heavily on a single relation between one property of the solar wind, speed at 1 AU, and one ratio of magnetic field strengths, the expansion factor, Equation \eqref{eq:test}. We compare the WSA relation to our results in Section 3.2.1, present a correlation between the Alfv\'en wave heating rate and the magnetic field strength in Section 3.2.2 and discuss important correlations between the magnetic field and temperatures in Section 3.2.3.
\subsubsection{Revisiting the WSA Model}
\begin{figure}
\includegraphics[width=0.45\textwidth]{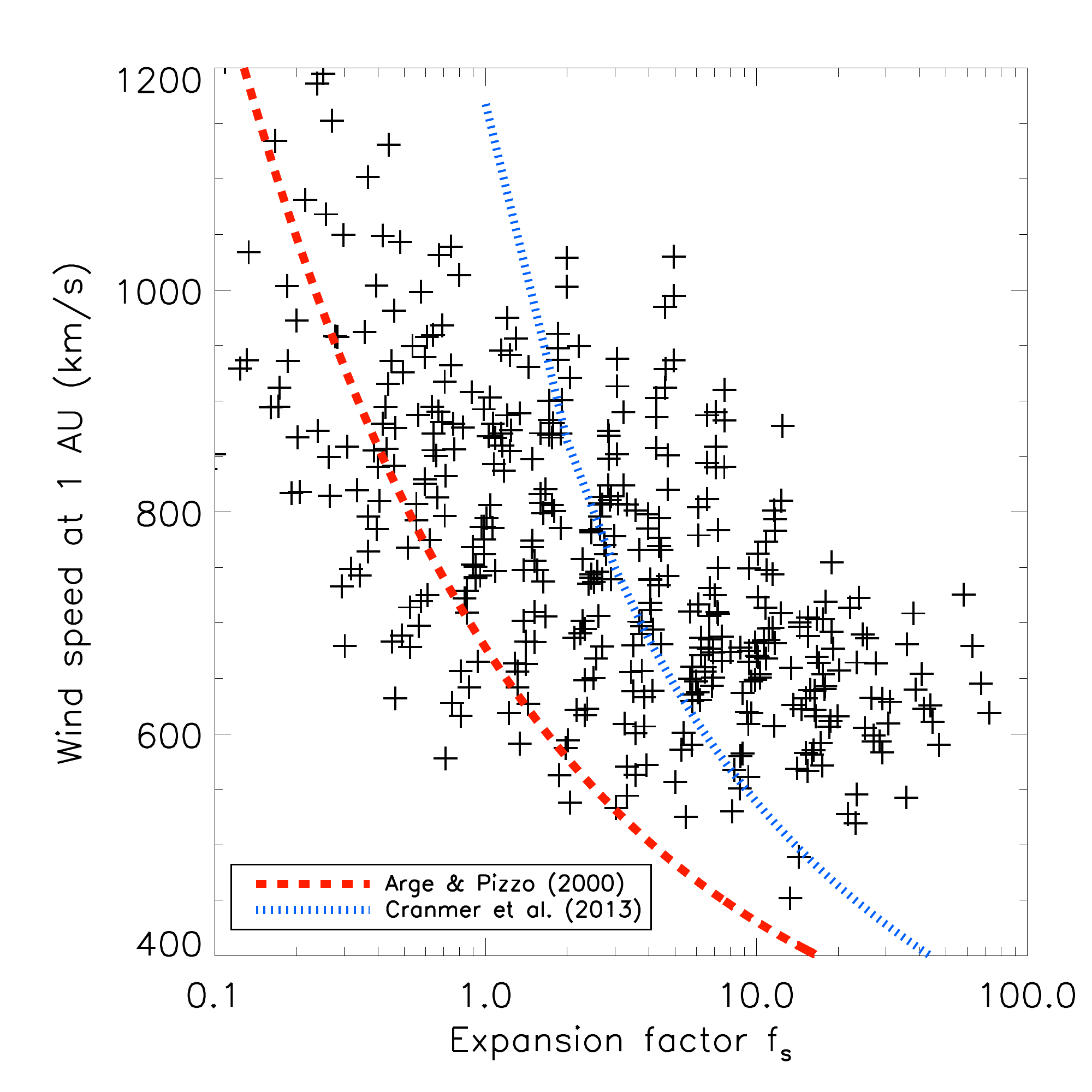}
\caption{Expansion factor anti-correlation is reproduced by ZEPHYR. The red dashed line shows the source surface velocity relation to expansion factor, Equation \eqref{eq:argepizzo}, from \cite{2000JGR...10510465A}, which is expected to be lower than the wind speed at 1 AU. The blue dotted line shows the concordance relation found by \cite{2013ApJ...767..125C}}.
\label{fig:zephyr_wsa}
\end{figure}

Our models follow the general anti-correlation of wind speed and expansion factor seen in observations \citep{1990ApJ...355..726W}, shown in Figure \ref{fig:zephyr_wsa}. There is nevertheless a large scatter around any given one-to-one relation between $u_{\infty}$ and $f_{s}$, which is highly reminiscent of the observed solar wind. That our models reproduce an observation-based relation is an important and successful test of the validity of ZEPHYR. The concordance relation found by \cite{2013ApJ...767..125C}, $u_{\infty} = 2300/(\ln{f_{s}} + 1.97)$, runs through the middle of the scatter for expansion factors greater than 2.0, just as it does for that paper.
\subsubsection{Alfv\'en Wave Heating Rate}
\begin{figure}
\includegraphics[width=0.45\textwidth]{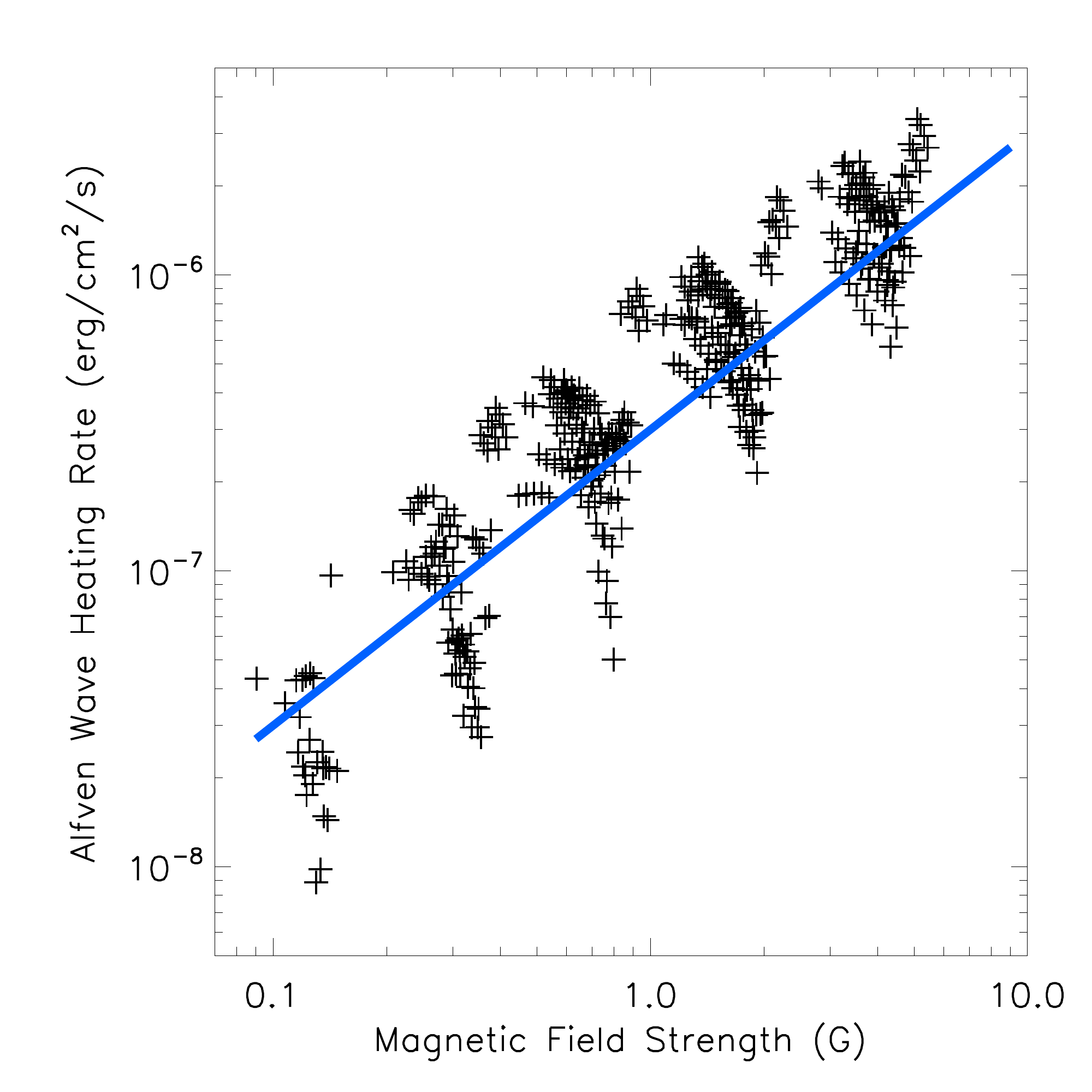}
\caption{Heating rate versus magnetic field strength at a height of 0.25 solar radii. At these low heights, this relation, shown with a solid line of slope = 1, is expected from turbulent damping (see Section 3.2.2).}
\label{fig:QvB}
\end{figure}

The turbulent heating rate by Alfv\'en waves can be written in terms of the Els\"asser variables $Z_{-}$ and $Z_{+}$ as the following:\begin{equation}Q_{A} = \rho \epsilon_{\rm turb}\frac{Z_{-}^{2}Z_{+} + Z_{+}^{2}Z_{-}}{4L_{\perp}}\end{equation} The effective turbulence correlation length $L_{\perp}$ is proportional to $B^{-1/2}$ and $\epsilon_{turb}$ is the turbulence efficiency \citep[see][]{2007ApJS..171..520C}. The ZEPHYR code iterates to find a value of $Q_{A}$ that is consistent with the time-steady conservation equations (Equation 3). For Alfv\'en waves at low heights where the solar wind speed is much smaller than the Alfv\'en speed, the Els\"asser variables are roughly proportional to $\rho^{-1/4}$. Putting this together with the thin flux-tube limit where the Alfv\'en speed is roughly constant, we show that Alfv\'en wave heating should produce, at low heights, the proportionality $Q_{A} \propto B$ \citep[see also][]{2009LRSP....6....3C}. Figure \ref{fig:QvB} shows this relation at a height of 0.25 solar radii. This also suggests that the magnetic field and temperature profiles should be reasonably well-correlated, and we will show in Section 4 that this is true.
\subsubsection{Predicted Temperatures}
\begin{figure}
\includegraphics[width=0.5\textwidth]{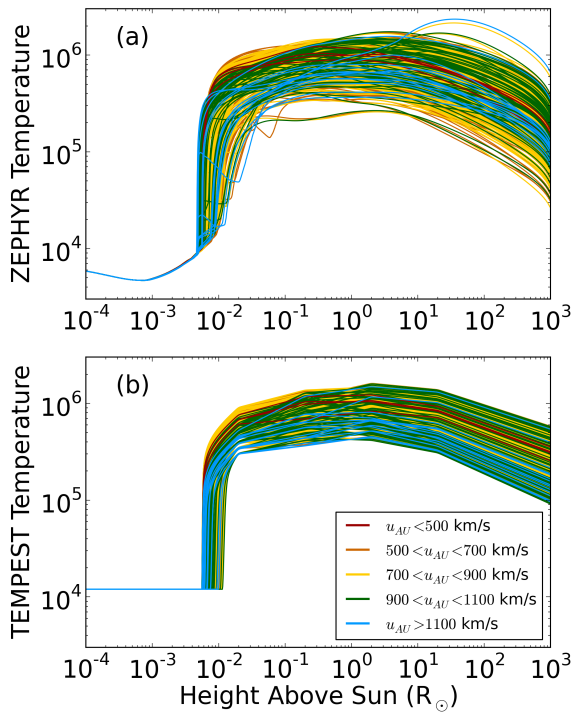}
\vspace{-12mm}
\caption{Comparison between (a) calculated temperature profiles from internal energy conservation in ZEPHYR and (b) the temperature profiles we set up for TEMPEST based on correlations with magnetic field strengths (see Appendix for more information).}
\label{fig:tempest_temps}
\vspace{2mm}
\end{figure}

\begin{figure}
\includegraphics[width=0.5\textwidth]{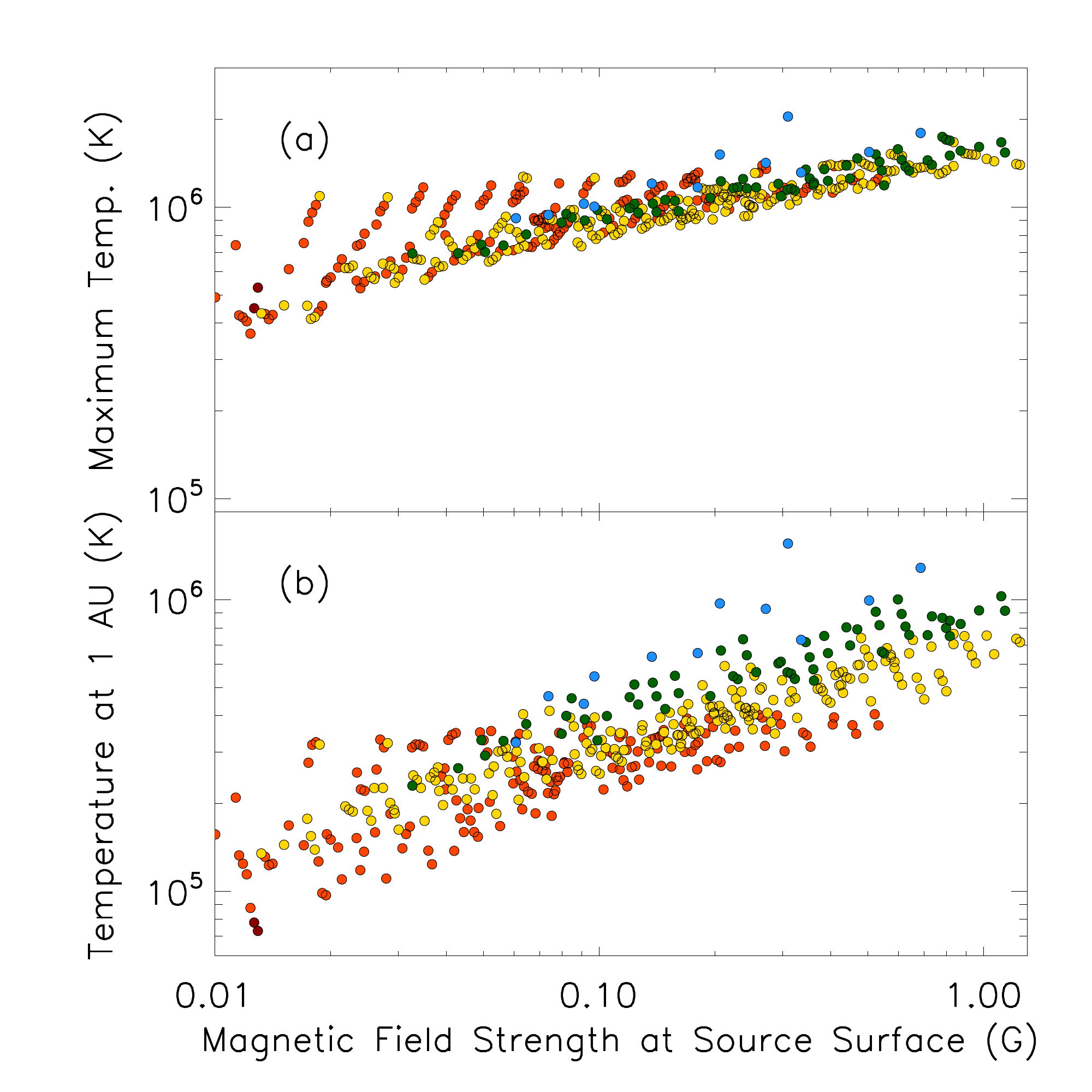}
\caption{Strong correlation of (a) the maximum temperature and (b) T at 1 AU with the magnetic field strength at the source surface. Color indicates outflow speed as in Figures \ref{fig:zt25} and \ref{fig:tempest_temps}.}
\label{fig:tempcorr}
\end{figure}

\begin{figure}
\includegraphics[width=0.45\textwidth]{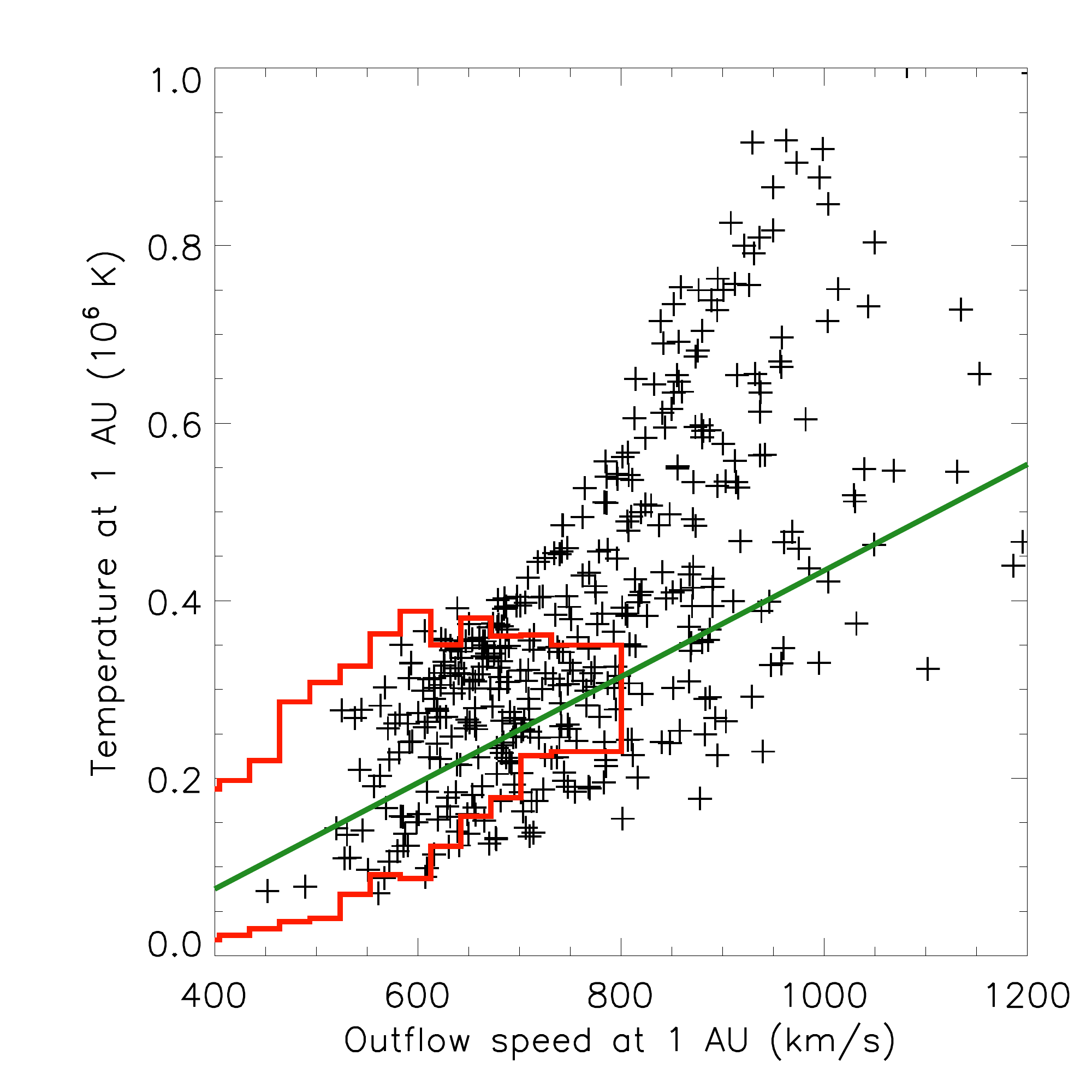}
\vspace{-2mm}\caption{Temperature and Wind Speed at 1 AU: red outline represents several decades of OMNI data and green line is the observationally derived linear fit of relation between proton temperatures and wind speed \citep{2012JGRA..117.9102E}.}
\label{fig:rpr_temp}
\end{figure}

ZEPHYR uses a simplified treatment of radiative transfer to compute the heating and cooling rates throughout the solar atmosphere. \cite{2007ApJS..171..520C} include terms for radiation, conduction, heating by Alfv\'en waves, and heating by acoustic waves. The photospheric base and lower chromosphere are considered optically thick and are dominated by continuum photons in local thermodynamic equilibrium that provide the majority of the heating and cooling. However, in the corona, the atmosphere is optically thin, where many spectral lines contribute to the overall radiative cooling. Further description of the internal energy conservation terms listed in Equation (3c) can be found in Section 3 of \cite{2007ApJS..171..520C}.\\
\indent The temperature profiles found for each flux tube model are presented in Figure \ref{fig:tempest_temps}. The blue models have speeds greater than 1100 km s$^{-1}$, and their temperature profiles peak higher than the mean height of maximum temperature. These models probably do not correspond to situations realized in the actual solar wind, but they are instructive as examples of the implications of extreme values of $B(r)$.\\
\indent Figure \ref{fig:tempcorr} shows illustrative correlations between the maximum temperature and the temperature at 1 AU with the magnetic field at $r = 2.5$ R$_{\odot}$. These are both very strong correlations (Pearson coefficients $R > 0.8$), and they can be used as an independent measure of the magnetic field near the source surface besides PFSS extrapolations from magnetogram data to test the overall validity of turbulence-driven models. If the measured solar wind exhibits a similar correlation between, e.g., temperature at 1 AU and the field strengths at a field line's extrapolated location at 2.5 R$_{\odot}$, this would provide additional evidence in favor of WTD-type models.\\
\indent Additionally, we show the relation between the temperature at 1 AU and the wind speed at 1 AU in Figure \ref{fig:rpr_temp}. We plot the linear fit between proton temperature and wind speed found by \cite{2012JGRA..117.9102E}, which is a good fit to models with wind speeds at or below 800 km s$^{-1}$. Models with higher wind speeds may be generated from slightly unphysical magnetic field profiles in our grid. We also plot the outline of the OMNI data set, which includes several decades of ACE/Wind data for proton temperatures and outflow speed. Our models populate the same region and spread for wind speeds between 550 and 700 km s$^{-1}$. We discuss the relative lack of slow solar wind results, i.e. $u \lesssim 400$ km s$^{-1}$, in Section 6.

\section{TEMPEST Development}
We developed The Efficient Modified-Parker-Equation-Solving Tool (TEMPEST) in Python, in order to provide the community with a fast and flexible tool that can be used as a whole or in parts due to its library-like structure. TEMPEST can predict the outflow speeds of the solar wind based only on the magnetic field profile of an open flux tube, which could be measured using PFSS extrapolations from magnetogram data. TEMPEST uses the modified Parker equation given in Equation \eqref{eq:parkereq}, but we neglect the small term proportional to $Q_{A}$ for simplicity. For a given form of the critical speed $u_{c}$ (see Section 4.1 and Section 4.2), the critical radius, $r_{c}$, is found as described in Section 3.1. At each critical point, the slope of the outflow must be found using L'H\^opital's Rule. Doing so, one finds \begin{equation}\label{eq:critslope}\left.\frac{du}{dr}\right|_{r=r_{c}} = \frac{1}{2}  \left[ \frac{du_{c}}{dr} \pm \sqrt{\left(\frac{du_{c}}{dr}\right)^{2} + 2\frac{d{\rm (RHS)}}{dr}}\right]\end{equation} 
where the postive sign gives the accelerating solution appropriate for the solar wind, and RHS is the right-hand side of Equation \eqref{eq:parkereq}. We emphasize that the actual wind-speed gradient $du/dr$ at the critical point is not the same as the gradient of the critical speed $du_{c}/dr$. Consider the simple case of an isothermal corona without wave pressure, in which $u_c$ is just a constant sound speed and $du_{c}/dr = 0$.  Even in this case, Equation \eqref{eq:critslope} gives nonzero solutions for $du/dr$ at the critical point: a positive value for the transonic wind and a negative value for the Bondi accretion solution.\\
\indent The magnetic field profile is the only user input to TEMPEST, and the temperature profile is set up using temperature-magnetic field correlations from ZEPHYR, as we do not include the energy conservation equation in TEMPEST. We considered the correlations between temperatures and magnetic field strengths at different heights, similar to the results shown in Figure \ref{fig:tempcorr}. We set the temperatures at evenly spaced heights $z_{T}$ in log-space, and at each height we sought to find the heights $z_{B}$ at which the variation of the magnetic field strength best correlates with the temperature at $z_{T}$. At $z_{T} = 0.02 $ R$_{\odot}$, the results from ZEPHYR give the best correlation with the magnetic field in the low chromosphere. At $z_{T} = 0.2 $ R$_{\odot}$, the temperature best correlates with the magnetic field at $z_{B} = 0.4 $ R$_{\odot}$. Since the temperature peaks around this middle height, the correlations reflect the fact that heat conducts away from the temperature maximum. At $z_{T} = $ 2, 20, and 200  R$_{\odot}$, the magnetic field near the source surface ($z_{B} \approx 2 - 3 $ R$_{\odot}$) provides the best correlation. We show the comparison between the temperature profiles of ZEPHYR and TEMPEST in Figure \ref{fig:tempest_temps} and provide the full equations in the Appendix.\\
\indent TEMPEST has two main methods of use. The first is a vastly less time-intensive mode we will refer to for the remainder of this paper as ``Miranda'' that solves for the outflow solution without including the wave pressure term. We outline this in Section 4.1. Miranda can run 200 models in under 60 seconds, making it a useful educational tool for showing how the magnetic field can affect the solar wind in a relative sense. It is important to note that the way that the temperature profiles are set up in TEMPEST means that Miranda already includes the effects of turbulent heating even though it does not have the wave pressure term, effectively separating the two main ways that Alfv\'en waves contribute to the acceleration of the solar wind. The second mode of TEMPEST use is the full outflow solver based on including both the gas and wave pressure terms. We will call this function ``Prospero'' for ease in reference, and we outline the additional steps that Prospero takes, after the inital solution is found using Miranda, in Section 4.2.
\subsection{Miranda: Without waves}
\begin{figure}
\includegraphics[width=0.45\textwidth]{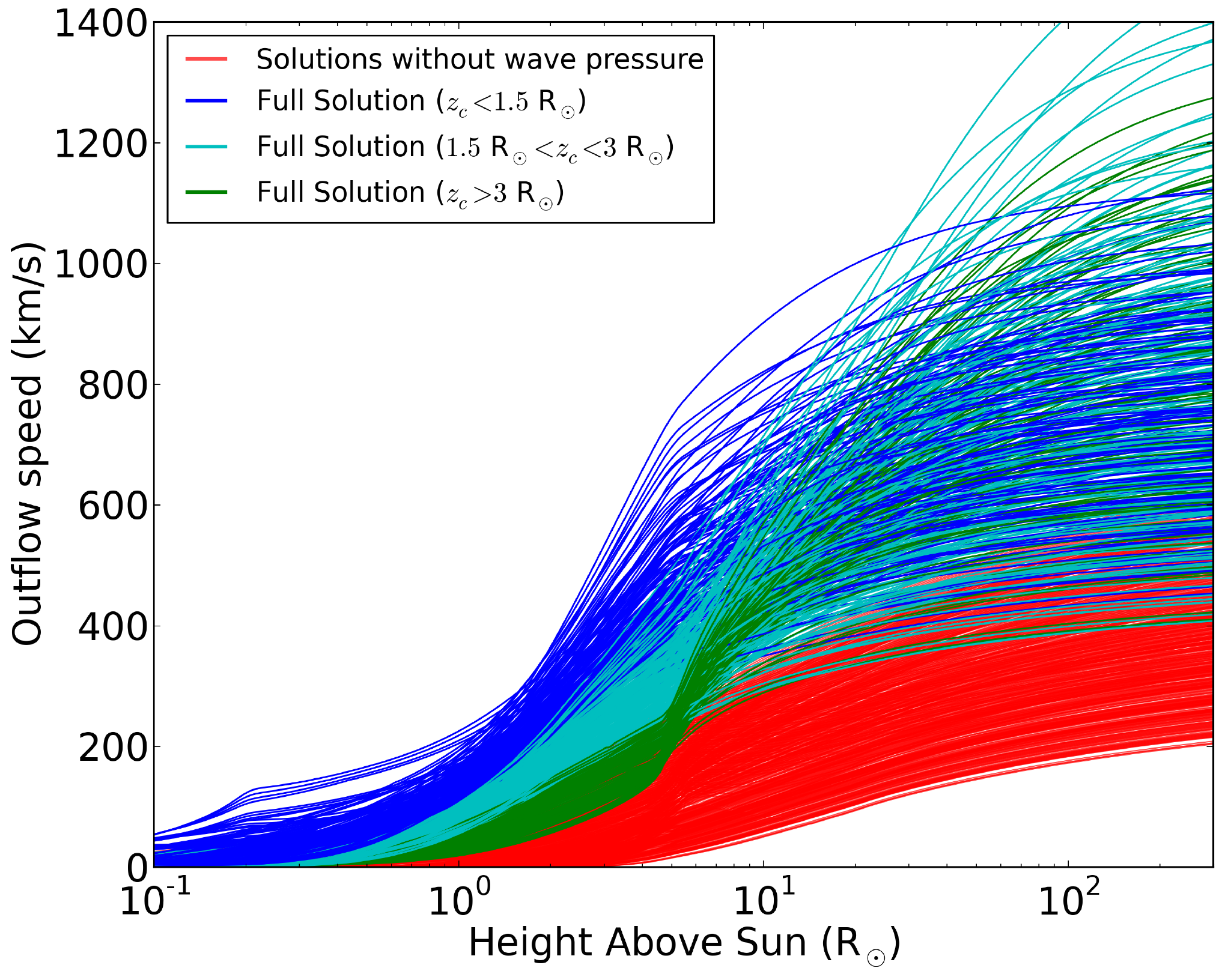}
\caption{Results from Prospero represent the full solution. The solutions from Miranda are shown in red. Color indicates bands of critical location height as shown in the legend. We plot only the 428 models from the grid that were well-converged in ZEPHYR, although TEMPEST iterates until all the models find a stable solution.}
\label{fig:prospero}
\end{figure}

The first step towards the full outflow solution requires calculating an initial estimate the outflow without waves in order to calculate the density profile (details in the following section). This first step, Miranda, solves Equation \eqref{eq:parkereq} with the terms for gravity, the magnetic field gradient, and the temperature gradient, where the critical speed $u_{c}$ is set to the isothermal sound speed, $a = (kT/m)^{1/2}$. The temperature profiles that TEMPEST uses already include the effects of turbulent heating, so we can effectively separate the two primary mechanisms by which Alfv\'en waves accelerate the solar wind: turbulent heating and wave pressure. The solution found using Miranda has only the first of these mechanisms included, and therefore will predict outflows at consistently lower speeds.\\
\indent With all terms in the Parker equation defined, we find the critical point as discussed in Section 3.1. Using the speed and radius of the correct critical point, we find the slope at the critical point using Equation \eqref{eq:critslope}. Once we have determined the critical point and slope, we use a 4th-order Runge-Kutta integrator to move away in both directions from this point.\\
\indent The results from TEMPEST without the wave pressure term are shown in red in Figure \ref{fig:prospero} and result in much lower speeds than ZEPHYR produced, which is to be expected, as the additional pressure from the waves appears to provide an important acceleration for the solar wind. The mean wind speed at 1 AU for the results from ZEPHYR is 776 km s$^{-1}$ (standard deviation is 197 km s$^{-1}$); the mean wind speed at 1 AU after running Miranda is only 357 km s$^{-1}$ with a standard deviation of 105 km s$^{-1}$. Therefore, we now look at the solutions to the full modified Parker equation used by Prospero.
\subsection{Prospero: Adding waves and damping}
\begin{figure}
\includegraphics[width=0.5\textwidth]{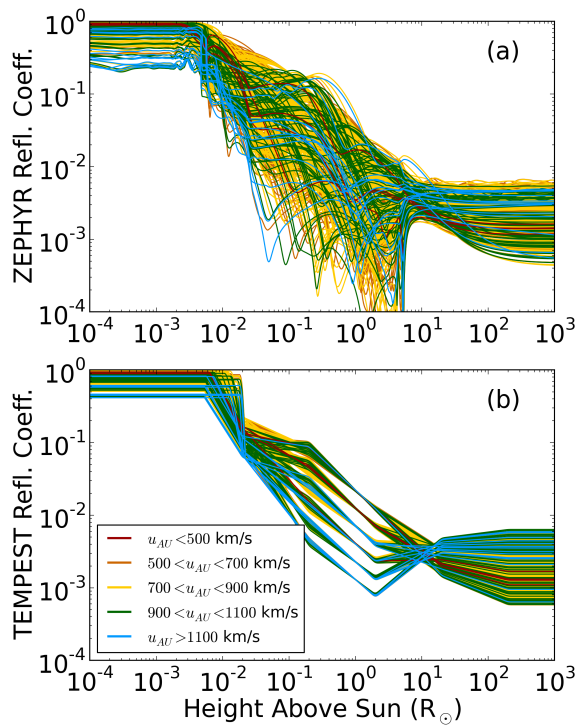}
\vspace{-12mm}
\caption{Comparison between (a) calculated reflection coefficients from ZEPHYR and (b) the defined reflection coefficients from TEMPEST (see Appendix for further details).}
\label{fig:refl}
\vspace{2mm}
\end{figure}

\begin{figure}
\includegraphics[width=0.45\textwidth]{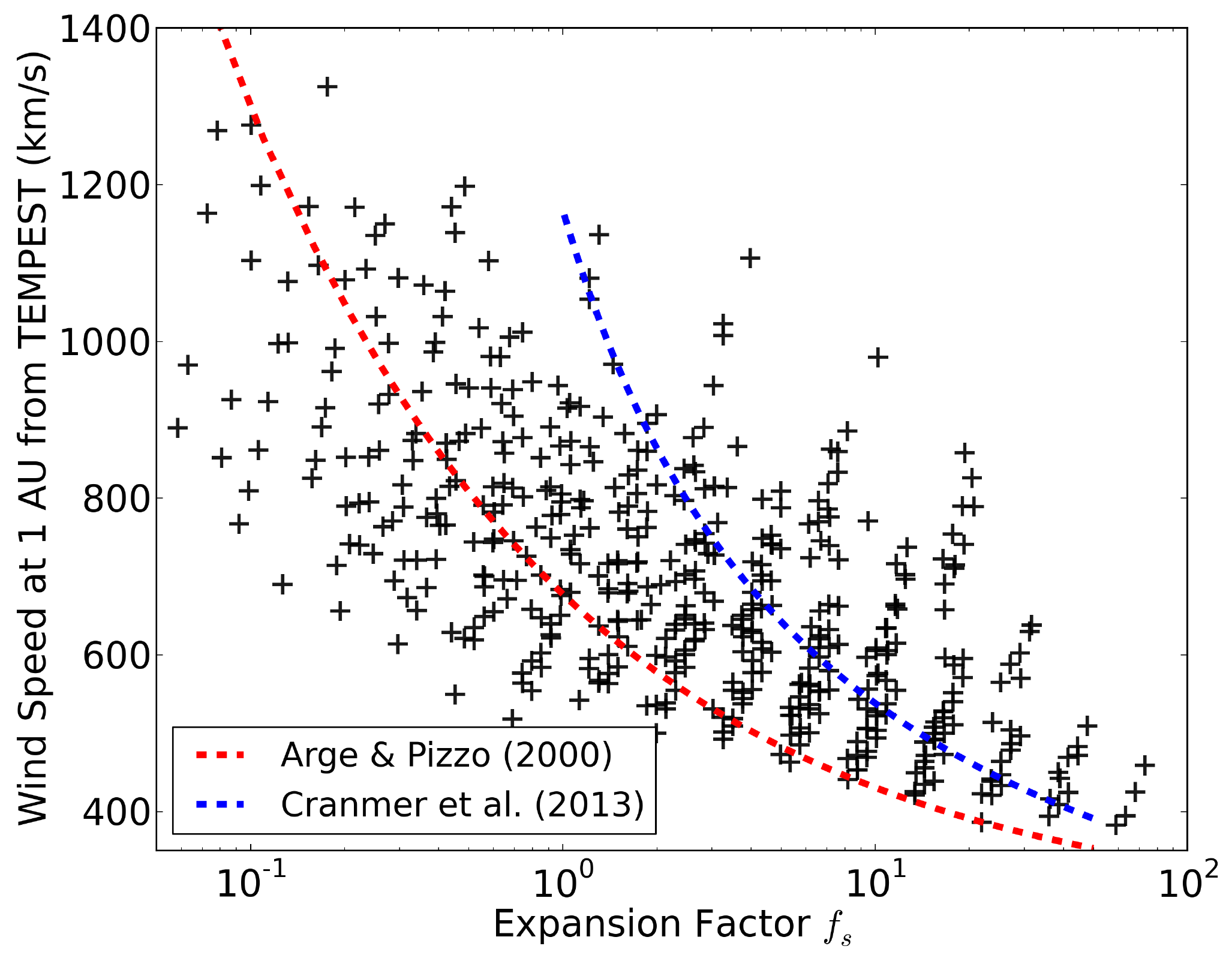}
\caption{Here we show a plot similar to Figure \ref{fig:zephyr_wsa}, now for results from TEMPEST. Again, the red dashed line is wind speed {\it at the source surface}, given by Equation \eqref{eq:argepizzo}, which should be lower than the wind speed at 1 AU for most of the models. The blue dashed line, as in Figure \ref{fig:zephyr_wsa}, is the concordance relation given by \cite{2013ApJ...767..125C}.}
\label{fig:tempest_wsa}
\end{figure}

Everything presented in the previous section remains the same for the full solution except for the form of the critical speed. With waves, we must use the full form, given by Equation \eqref{eq:ucrit}. The mass density, $\rho$, is determined by using the outflow solution and the enforcement of mass flux conservation:
\begin{equation} \frac{u(r)\rho(r)}{B(r)} = \frac{u_{TR}\rho_{TR}}{B_{TR}}\end{equation} We set the transition region density based on a correlation with the transition region height that we found in the collection of ZEPHYR models, \begin{equation}\log(\rho_{TR}) = -21.904-3.349\log(z_{TR}),\end{equation}
where $\rho_{TR}$ is specified in g cm$^{-3}$ and $z_{TR}$ is given in solar radii. Although the pressure scale height differs at the transition region between ZEPHYR and TEMPEST (see Figure \ref{fig:tempest_temps}), we found the uncertainties produced by this assumption were small. Our initial version of TEMPEST used a constant value of this density, taken from the average of the ZEPHYR model results, and did not create significant additional disagreement between the ZEPHYR and TEMPEST results.\\ \indent We use damped wave action conservation to determine the Alfv\'en energy density, $U_{A}$. We start with a simplified wave action conservation equation to find the evolution of $U_{A}$:\begin{equation}\label{eq:waveact1}\frac{\partial}{\partial t}\left(\frac{U_{A}}{\omega'}\right) + \frac{1}{A}\frac{\partial}{\partial r}\left(\frac{[u+V_{A}]AU_{A}}{\omega'}\right) = -\frac{Q_{A}}{\omega'}\end{equation}
\citep[see e.g.][]{1977ApJ...215..942J, 2007ApJS..171..520C}. Because we are working with the steady-state solution to the Parker equation, we are able to neglect the time derivative. The Doppler-shifted frequency in the solar wind frame, $\omega'$, can be written as $\omega' = \omega V_{A}/(u+V_{A})$ where $\omega$ is a constant and may be factored out. The exact expression for the heating rate $Q_{A}$ depends on the Els\"asser variables $Z_{+}$ and $Z_{-}$, but we approximate it following \cite{2011ApJ...741...54C} as \begin{equation}\label{eq:qa}Q_{A} = \frac{\widetilde{\alpha}}{L_{\perp}}\rho v_{\perp}^{3}\end{equation} where the efficiency factor $\widetilde{\alpha} = 2\epsilon_{turb}R(1+R)/[(1+R^{2})^{3/2}]$. For TEMPEST, we define a simplified radial profile for the reflection coefficient $R$ based on correlations with the magnetic field strength in ZEPHYR (see Figure \ref{fig:refl} and Appendix). We set the correlation length at the base of the photosphere, $L_{\perp\odot}$, to 75 km \citep{2005ApJS..156..265C,2007ApJS..171..520C} and use the relation $L_{\perp} \propto B^{-1/2}$ for other heights (see Section 3.2.2). Combining Equations \eqref{eq:waveact1} and \eqref{eq:qa} using the approximations mentioned and including the conservation of magnetic flux, we define the wave action as \begin{equation}S \equiv \frac{(u+V_{A})^{2}\rho v_{\perp}^{2}}{BV_{A}}\end{equation} such that the wave action conservation equation can now be written as the following:\begin{equation}\frac{dS}{dr} = -S^{3/2}\left(\frac{\widetilde{\alpha}}{L_{\perp}(u+V_{A})^{2}}\right)\sqrt{\frac{BV_{A}}{\rho}}\end{equation} TEMPEST then integrates using a Runge-Kutta method to solve for $S(r)$ and uses a value of this constant at the photospheric base, $S_{base} = 5 \times 10^{4} $ erg/cm$^{2}$/s/G (which was assumed for each of the ZEPHYR models), to obtain the Alfv\'en energy density needed by the full form of the critical speed, such that:\begin{equation}U_{A}(r) \equiv \rho v_{\perp}^{2} = \frac{S(r)B(r)V_{A}(r)}{(u(r)+V_{A}(r))^{2}}\end{equation}
\indent To converge to a stable solution, Prospero must iterate several times. We use undercorrection to make steps towards the correct outflow solution, such that $u_{next} = u_{previous}^{0.9}*u_{current}^{0.1}$. The first iteration uses the results from Miranda as $u_{previous}$ and an initial run of Prospero using this outflow solution to provide $u_{current}$, and subsequent runs use neighboring iterations of Prospero to give the outflow solution to provide to the next iteration. The converged results from Prospero are presented in Figure \ref{fig:prospero}. The outflow speeds are between 400 and 1400 km s$^{-1}$ at 1 AU, consistent with observations (mean: 794 km s$^{-1}$, standard deviation: 199 km s$^{-1}$). For this figure, we have broken the models up by color according to the height of the critical point in order to highlight the relation between critical point height and asymptotic wind speed.\\
\indent A key result from the TEMPEST results is the recovery of a WSA-type relation, Equation \eqref{eq:argepizzo}. In Figure \ref{fig:tempest_wsa}, we plot the relation along with the results from the 428 well-converged models in our grid. The Arge \& Pizzo relation predicts the wind speed at the source surface based on the expansion factor (Equation \eqref{eq:test}). The relation should act as a lower bound for the wind speed predicted at 1 AU, since there is further acceleration above 2.5 $R_{\odot}$. This is exactly what we see for the slower wind speeds, which is to be expected for a relation calibrated at the equator, which rarely sees the highest speed wind streams. Both the ZEPHYR and TEMPEST models naturally produce a substantial spread around the mean WSA-type relation, highlighting the need the take the full magnetic field profile into account.\\

\section{Code Comparisons}
\begin{figure}
\includegraphics[width=0.5\textwidth]{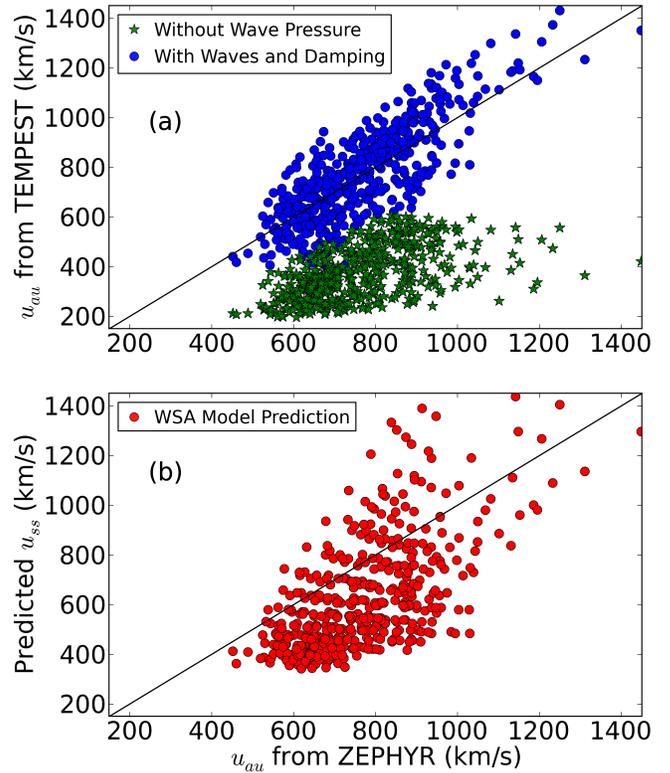}
\vspace{-6mm}\caption{(a) Wind speeds determined by TEMPEST compared to those found by ZEPHYR. The black line represents agreement; Models that reached a steady state solution in ZEPHYR are highlighted in blue for Prospero results and green for the initial Miranda solutions. (b) Predictions using Equation \eqref{eq:argepizzo} compared to ZEPHYR results. We expect the overall lower speeds, but the scatter is much greater when using WSA to make predictions instead of TEMPEST.}
\label{fig:zt}
\end{figure}

In Figure \ref{fig:zt}, we show directly the wind speeds determined by both modes of TEMPEST and by ZEPHYR. The solutions found by Miranda have a minimum speed at 1 AU around 200 km s$^{-1}$, similar to the observed lower limit of in situ measurements. Figure \ref{fig:zt}a highlights the two discrete ways in which Alfv\'en waves contribute to the acceleration of the solar wind. It is important to note that the scatter in comparing ZEPHYR and a WSA prediction (Figure \ref{fig:zt}b) is much greater than the scatter in the ZEPHYR-TEMPEST comparison, due to the magnetic variability ignored by using only the expansion factor to describe the geometry. The root-mean-square (RMS) difference in the ZEPHYR-TEMPEST comparison (the blue points in Figure  \ref{fig:zt}a) is 115 km s$^{-1}$, while the RMS difference in the ZEPHYR-WSA comparison (the red points in Figure \ref{fig:zt}b) is 228 km s$^{-1}$. The average percent difference between the computed speeds for ZEPHYR and those of the full mode of TEMPEST for each model is just under 14\%. We also ran the same 628 models through a version of TEMPEST that directly reads in the temperature and reflection coefficient profiles, and the percent difference was just below 12\%. We discuss other possible improvements to TEMPEST to lower this scatter in Section 6. These numbers indicate that TEMPEST, while it makes many simplifying assumptions, is a more consistent predictor of wind speeds than the traditional observationally-derived WSA approach, which does not specify any particular choice of the underlying physics that accelerates the wind. There does exist the possibility that WSA predictions better match observations than either TEMPEST or ZEPHYR, and our next steps will be to use the completed TEMPEST code, in combination with magnetic extrapolations of the coronal field, to predict solar wind properties for specific time periods and compare them with in situ measurements.\\
\indent Another important distinction between these two codes is CPU run-time. TEMPEST runs over forty times faster than ZEPHYR because it makes many simplifying assumptions. We intend to take advantage of the ease of parallel processing in Python to improve this speed increase further in future versions of the code.

\section{Discussion}
We have used WTD models to heat the corona through dissipation of heat by turbulent cascade and accelerate the wind through increased gas pressure and additional wave pressure effects. Our primary goal for this project is to improve empirical forecasting techniques for the steady-state solar wind. As we have shown, the community often relies on WSA modeling, based on a single parameter of the magnetic field expansion in open flux tubes. Even with the advances of combining MHD simulations as the WSA-ENLIL model, comparisons between predictions and observations make it clear that further improvements are still necessary. An important point to make is that extrapolations from magnetograms show that many flux tube magnetic field strengths do not all monotonically decrease, so two models with identical expansion factors could result in rather different structures. We anticipate that TEMPEST could easily be incorporated within an existing framework to couple it with a full MHD simulation above the source surface.\\ \indent The first code we discuss, ZEPHYR, has been shown to correspond well with observations of coronal holes and other magnetic structures in the corona. We investigate here the results of a grid of models that spans the entire range of observed flux tube strengths throughout several solar cycles to test the full parameter space of all possible open magnetic field profiles.\\ \indent ZEPHYR also provides us with temperature-magnetic field correlations that help to take out much of the computation time for a stand-alone code, TEMPEST, that solves the momentum conservation equation for the outflow solution of the solar wind based on a magnetic field profile. The solar physics community has come a long way since Parker's spherically symmetric, isothermal corona, but the groundwork laid by this early theory is still fully applicable.\\ \indent The special case presented by pseudostreamers is an ongoing area of our analysis. Pseudostreamers do not contribute to the heliospheric current sheet and they seem to be a source of the slow solar wind. The community does not fully understand the differences in the physical properties of the solar wind that may emanate from pseudostreamers and helmet streamers, although observational evidence suggests the slow wind is generated from these areas or the edges of coronal holes. Our results from both codes do not currently recreate the bimodal distribution of wind speeds observed at 1 AU. It is unclear whether this is due to the inclusion of many unphysical flux tube models or because the fast wind and the slow wind are generated by {\it different} physical mechanisms.\\ \indent One slightly troubling feature of the ZEPHYR and TEMPEST model results is a relative paucity of truly ``slow'' wind streams ($u \lesssim 350$ km s$^{-1}$) in comparison to the observed solar wind.  However, \cite{2011JGRA..116.3106M} showed that many of the slowest wind streams at 1 AU were the result of gradual deceleration due to stream interactions between 0.1 and 1 AU.  Similarly, \cite{2013ApJ...767..125C} found that ZEPHYR models of near-equatorial quiet-Sun stream lines exhibited a realistic distribution of slow speeds at 0.1 AU, but they exhibited roughly 150 km s$^{-1}$ of extra acceleration out to 1 AU when modeled in ZEPHYR without stream interactions.  Clearly, taking account of the development of corotating interaction regions and other stream-stream effects is key to producing more realistic predictions at 1 AU.\\ \indent Another important avenue of future work will be to compare predictions of wind speeds from TEMPEST with in situ measurements, when the results from ZEPHYR and TEMPEST agree to a greater extent. We are already able to reproduce well-known correlations and linear fits from observations, but accurate forecasting is our goal. Other ways in which forecasting efforts can be improved that TEMPEST does not address include better lower boundary conditions on and coronal extrapolation of $\vec{B}$, moving from 1D to a higher dimensional code, and including kinetic effects of a multi-fluid model ($T_{p} \ne T_{e}$, $T_{\parallel} \ne T_{\perp}$).\\\indent Space weather is dominated by both coronal mass ejections (CMEs) and high-speed wind streams. The latter is well-modeled by the codes presented in this paper, and these high-speed streams produce a greatly increased electron flux in the Earth's magnetosphere, which can lead to satellite disruptions and power-grid failure \citep{2011A&A...526A..20V}. Understanding the Sun's effect on the heliosphere is also important for the study of other stars, especially in the ongoing search for an Earth analog. The Sun is an indespensible laboratory for understanding stellar physics due to the plethora of observations available. The modeling we have done in this project marks an important step toward full understanding of the coronal heating problem and identifying sources of solar wind acceleration.

\section*{Acknowledgments}
This material is based upon work supported by the National Science Foundation Graduate Research Fellowship under Grant No. DGE-1144152 and by the NSF SHINE program under Grant No. AGS-1259519.  The authors obtained the OMNI solar wind data from the GSFC/SPDF OMNIWeb interface and thank David McComas and Ruth Skoug (ACE/SWEPAM) and Charles Smith and Norman Ness (ACE/MAG) for providing the majority of the OMNI measurements used in this paper. L.N.W. also thanks the Harvard Astronomy Department for the student travel grant and Loomis fund.

\appendix
\section{APPENDIX: Temperature and Reflection Coefficient Profiles}
One of the primary differences between the ZEPHYR and TEMPEST codes is the way in which internal energy conservation is handled. ZEPHYR finds a self-consistent solution for the equations of mass, momentum, and energy conservation, including the physical processes of Alfv\'en wave-driven turbulent heating. TEMPEST is meant to be a stand-alone code that runs faster by making reasonable assumptions about these processes. To do so, we use correlations with the magnetic field to set up the temperature and Alfv\'en wave reflection coefficient profiles. We describe this process here.\\ \indent The temperature profile can be described by a relatively constant-temperature chromosphere that extends to the transition region height, $z_{TR}$, a sharp rise to the location of the temperature peak, $z_{max}$, followed by a continued gradual decrease. We found that $z_{TR}$ was best correlated with the strength of the magnetic field at $z_{B} = 2.0$ R$_{\odot}$, and this fit (with Pearson correlation coefficient $\mathcal{R} = -0.29$) is given by \begin{equation}z_{TR} = 0.0057 + \left(\frac{7 \times 10^{-6}}{B(2.0 ~R_{\odot})^{1.3}}\right) R_{\odot}\end{equation}.
\begin{subequations}
We then chose evenly-spaced heights $z_{T}$ in log-space to set the temperature profile according to the following set of linear fits in log-log space with the location along the magnetic field profile that best correlated (Pearson coefficients $\mathcal{R}$ given for each):
\begin{align}
\log_{10}(T(0.02 {\rm ~R}_{\odot})) = 5.554 + 0.1646\log_{10}(B(0.00314 {\rm ~R}_{\odot})) ~~~~~\mathcal{R} = 0.51 \\
\log_{10}(T(0.2 {\rm ~R}_{\odot})) = 5.967 + 0.2054\log_{10}(B(0.4206 {\rm ~R}_{\odot})) ~~~~~\mathcal{R} = 0.83 \\
\log_{10}(T(2.0 {\rm ~R}_{\odot})) = 6.228 + 0.2660\log_{10}(B(2.0 {\rm ~R}_{\odot})) ~~~~~\mathcal{R} = 0.91 \\
\log_{10}(T(20 {\rm ~R}_{\odot})) = 5.967 + 0.2054\log_{10}(B(3.0 {\rm ~R}_{\odot})) ~~~~~\mathcal{R} = 0.87 \\
\log_{10}(T(200 {\rm ~R}_{\odot})) = 5.967 + 0.2054\log_{10}(B(3.0 {\rm ~R}_{\odot})) ~~~~~\mathcal{R} = 0.84
\end{align}
\end{subequations}
\begin{subequations}
We looked for correlations in the residuals of each of these fits, and found well-correlated ($\mathcal{R} > 0.45$) terms for the first two heights, at $z_{T} = 0.02$ R$_{\odot}$ and $0.2$ R$_{\odot}$. We added the following terms to the $\log(T)$ estimates given above in order to improve the overall correlation,
\begin{align}
\log_{10}(T_{resid}(0.02 {\rm ~R}_{\odot})) = 0.0559 + 0.13985\log_{10}(B(0.662 {\rm ~R}_{\odot}))\\
\log_{10}(T_{resid}(0.2 {\rm ~R}_{\odot})) = -0.0424 + 0.09285\log_{10}(B(0.0144 {\rm ~R}_{\odot})).
\end{align}
\end{subequations}
After adding these terms, the correlation coefficients improved to $\mathcal{R} = 0.78$ at $z_{T} = 0.02{\rm ~R}_{\odot}$ and $\mathcal{R} = 0.94$ at $z_{T} = 0.2{\rm ~R}_{\odot}$. With all of these fitted values, we then constructed the temperature profile with the following continuous piecewise function, where the chromosphere is at a constant temperature $T_{TR} = 1.2 \times 10^{4}$ K and $aT \equiv log_{10}(T)$:
\small
\[T(z) = \left\{
  \begin{array}{lr}
	T_{TR} & : z \le z_{TR}\\
	\left[T_{TR}^{3.5} + \left(\frac{T(0.02 {\rm ~R}_{\odot})^{3.5} - T_{TR}^{3.5}}{(0.02 {\rm ~R}_{\odot})^{2} - z_{TR}^{2}}\right)(z^{2} - z_{TR}^{2})\right]^{2/7} & : z_{TR} < z \le 0.02 {\rm ~R}_{\odot}\\
	10^{x}, x = \left(aT(0.02 {\rm ~R}_{\odot}) + \frac{aT(0.2 {\rm ~R}_{\odot}) - aT(0.02 {\rm ~R}_{\odot})}{\log_{10}(0.2) - \log_{10}(0.02) }(\log_{10}(z) + 1.7)\right) & : 0.02 {\rm ~R}_{\odot} < z \le 0.2 {\rm ~R}_{\odot}\\
	10^{x}, x = \left(aT(0.2 {\rm ~R}_{\odot}) + \frac{aT(2.0 {\rm ~R}_{\odot}) - aT(0.2 {\rm ~R}_{\odot})}{\log_{10}(2.0) - \log_{10}(0.2) }(\log_{10}(z) + 0.7)\right) & : 0.2 {\rm ~R}_{\odot} < z \le 2.0 {\rm ~R}_{\odot}\\
	10^{x}, x = \left(aT(2.0 {\rm ~R}_{\odot}) + \frac{aT(20 {\rm ~R}_{\odot}) - aT(2.0 {\rm ~R}_{\odot})}{\log_{10}(20) - \log_{10}(2.0) }(\log_{10}(z) - 0.3)\right) & : 2.0 {\rm ~R}_{\odot} < z \le 20 {\rm ~R}_{\odot}\\
	10^{x}, x = \left(aT(20 {\rm ~R}_{\odot}) + \frac{aT(200 {\rm ~R}_{\odot}) - aT(20 {\rm ~R}_{\odot})}{\log_{10}(200) - \log_{10}(20) }(\log_{10}(z) - 1.3)\right) & : z > 20 {\rm ~R}_{\odot}
  \end{array}
\right.
\]\\
\normalsize
\indent We proceeded with a similar method to create the Alfv\'en wave reflection coefficient $R$ used in TEMPEST. For the same evenly spaced in log-space heights ($0.02,~0.2,~2.0,~20,~200$ R$_{\odot}$), we found linear fits between $\log_{10}(B)$ and $\log_{10}(R)$. They are (with Pearson coefficients $\mathcal{R}$):
\begin{subequations}
\begin{align}
\log_{10}(R(0.02 {\rm ~R}_{\odot})) = -1.081 + 0.3108\log_{10}(B(0.011 {\rm ~R}_{\odot})) ~~~~~\mathcal{R} =  0.64\\
\log_{10}(R(0.2 {\rm ~R}_{\odot})) = -1.293 + 0.6476\log_{10}(B(0.573 {\rm ~R}_{\odot})) ~~~~~\mathcal{R} =  -0.20\\
\log_{10}(R(2.0 {\rm ~R}_{\odot})) = -2.238 + 0.0601\log_{10}(B(0.0.315 {\rm ~R}_{\odot})) ~~~~~\mathcal{R} = 0.70 \\
\log_{10}(R(20 {\rm ~R}_{\odot})) = -2.940 - 0.2576\log_{10}(B(3.0 {\rm ~R}_{\odot})) ~~~~~\mathcal{R} = -0.27 \\
\log_{10}(R(200 {\rm ~R}_{\odot})) = -3.404 - 0.4961\log_{10}(B(3.0 {\rm ~R}_{\odot})) ~~~~~\mathcal{R} = -0.38
\end{align}
\end{subequations}
It is important to note tht the correlations are not as strong for this set of fits as they were for the temperature profile. However, we found no additional strong correlations in the residuals, and the effect due to the difference in the reflection coefficient between ZEPHYR and TEMPEST is small. Finally, with these fits we defined the following continuous piecewise function, $aR \equiv \log_{10}(R)$:
\small
\[R(z) = \left\{
  \begin{array}{lr}
	\frac{B(0.00975 {\rm ~R}_{\odot})}{0.7 + B(0.00975 {\rm ~R}_{\odot})} & : z \le z_{TR}\\
	10^{x}, x=\left(   aR(z_{TR}) + \frac{aR(0.02 {\rm ~R}_{\odot}) - aT(z_{TR})}{\log_{10}(0.02) - \log_{10}(z_{TR}) }(\log_{10}(z) - \log_{10}(z_{TR}))    \right) & : z_{TR} < z \le 0.02 {\rm ~R}_{\odot}\\
	10^{x}, x=\left(aR(0.02 {\rm ~R}_{\odot}) + \frac{aR(0.2 {\rm ~R}_{\odot}) - aT(0.02 {\rm ~R}_{\odot})}{\log_{10}(0.2) - \log_{10}(0.02) }(\log_{10}(z) + 1.7)\right) & : 0.02 {\rm ~R}_{\odot} < z \le 0.2 {\rm ~R}_{\odot}\\
	10^{x}, x=\left(aR(0.2 {\rm ~R}_{\odot}) + \frac{aR(2.0 {\rm ~R}_{\odot}) - aT(0.2 {\rm ~R}_{\odot})}{\log_{10}(2.0) - \log_{10}(0.2) }(\log_{10}(z) + 0.7)\right) & : 0.2 {\rm ~R}_{\odot}< z \le 2.0 {\rm ~R}_{\odot}\\
	10^{x}, x=\left(aR(2.0 {\rm ~R}_{\odot}) + \frac{aR(20 {\rm ~R}_{\odot}) - aT(2.0 {\rm ~R}_{\odot})}{\log_{10}(20) - \log_{10}(2.0) }(\log_{10}(z) - 0.3)\right) & : 2.0 {\rm ~R}_{\odot}< z \le 20 {\rm ~R}_{\odot} \\
	10^{x}, x=\left(aR(20 {\rm ~R}_{\odot}) + \frac{aR(200 {\rm ~R}_{\odot}) - aT(20 {\rm ~R}_{\odot})}{\log_{10}(200) - \log_{10}(20) }(\log_{10}(z) - 1.3)\right) & : 20 {\rm ~R}_{\odot}< z \le 200 {\rm ~R}_{\odot}\\
R(200 {\rm ~R}_{\odot}) & : z > 200 {\rm ~R}_{\odot}\\
  \end{array}
\right.
\]\\
\normalsize
The final step we followed to set up both the temperature and reflection coefficient profiles was to smooth each of the piecewise functions with a Bartlett window $w(x)$ of width $N = 15$, where \begin{equation}w(x) = \frac{2}{N-1}\left(\frac{N-1}{2} - \left|x - \frac{N-1}{2}\right|\right).\end{equation} These final profiles are presented in Figures \ref{fig:tempest_temps}b and \ref{fig:refl}b.

\bibliographystyle{apj}
\end{document}